# From Micro-Cognition to Self-Construction:

# A Four-Layer Integrative Review of Psychological Theories in HCI

**Xiaohe Bie[1], Bo Wang[2,1]*, Xinyu Long[1]**

**1. Mental Health Center of Tianjin University, School of Education, Tianjin University, Tianjin 300350, China;**

**biexiaohe@tju.edu.cn, longxin_yu112@tju.edu.cn**

**2. School of Computer Science and Technology, Tianjin University, Tianjin 300350, China; bo_wang@tju.edu.cn**

**Abstract:**

Human-computer interaction (HCI) is undergoing a paradigm shift from "tool use" toward "partnership" and even "mind symbiosis," with psychology evolving from a supplementary explanatory tool to a core pillar shaping interaction paradigms and long-term relationships. This paper systematically reviews relevant research and proposes a four-layer integrative framework comprising the Micro-cognitive, Meso-affective, Macro-social, and Self-constructive layers. The framework reveals that: the cognitive layer constitutes the foundation of interaction, the affective layer drives relational engagement, the social layer regulates trust and behavior through norms, and the self-constructive layer points to the ultimate goal of human-machine symbiosis. These four layers follow a progressive logic of "foundation--mediation--context--goal." The review further argues that psychology has shifted from "post-hoc evaluation" to a "proactive design paradigm," with direct implications for emerging scenarios such as generative AI and embodied agents. Concurrently, this paper identifies core challenges including the nature of affect, agency erosion, and ethical risks, and proposes future directions such as bidirectional theory of mind and personalized co-evolution. This framework offers a systematic theoretical perspective for understanding complex human-machine relationships.

## 1 Introduction

The field of human-computer interaction (HCI) is undergoing a profound paradigm shift. At its core, this transition moves away from viewing technology as a passive "tool" and toward regarding it as an active, socially intelligent "partner." This transformation demands that HCI research and design move beyond traditional usability engineering and enter a more complex, dynamic domain--one in which psychological theories are no longer merely auxiliary instruments for explaining phenomena or optimizing interfaces, but rather serve as the theoretical foundation for proactively shaping interaction paradigms (Norman, 2010).

To chart the core developments in this field, this paper first outlines three major waves of HCI paradigm evolution, the corresponding three-stage transformation of psychology's role, and the four-layer integrative framework proposed herein. These three dimensions are not isolated; they exhibit deep internal mapping and logical correspondence, collectively forming a comprehensive picture of the restructuring of human-machine relationships, as illustrated in Figure 1.1.

* Corresponding Author

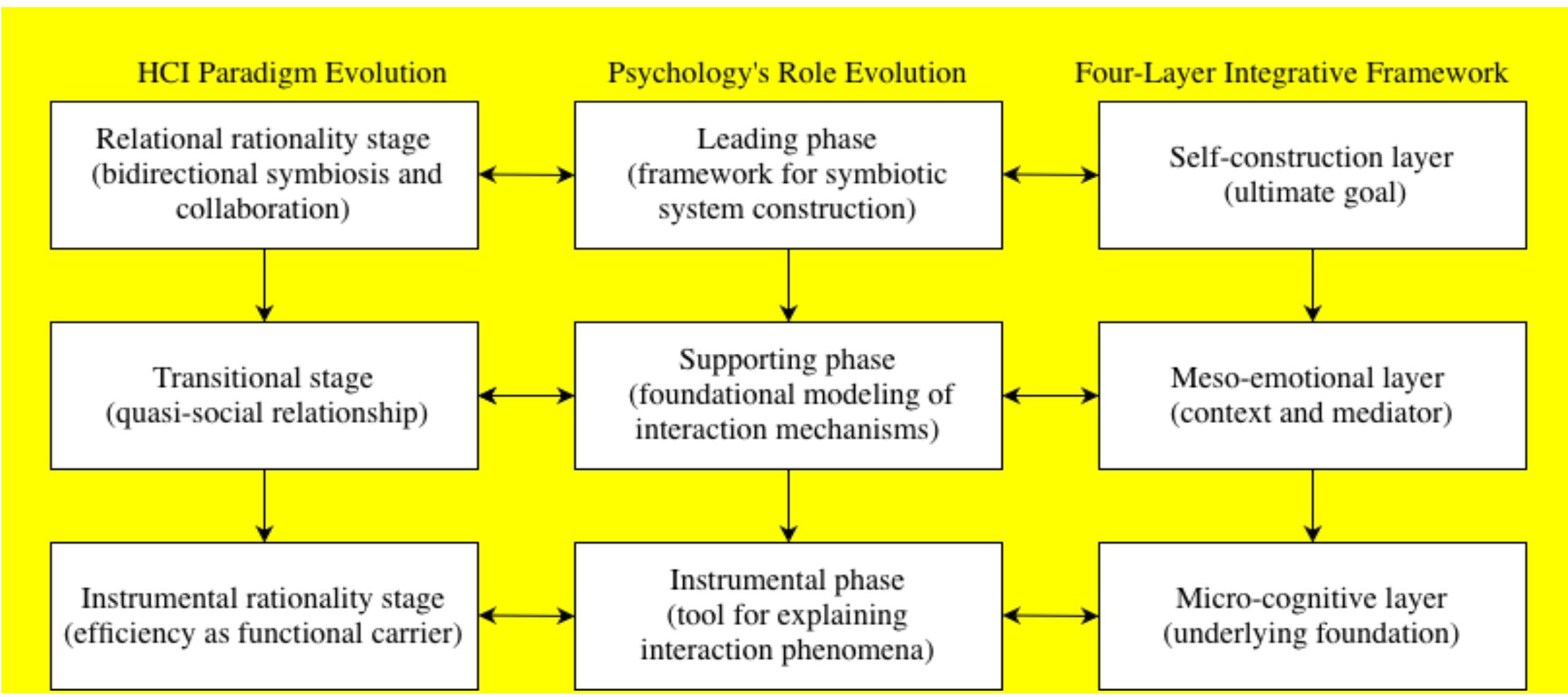

Figure 1.1 Correspondence between HCI Paradigm Evolution, Psychology's Role Evolution, and the Four-Layer Integrative Framework

Several review works have recently emerged that attempt to synthesize the landscape of human-machine relationships. For instance, some adopt a computer science perspective, focusing on the technical classification and algorithmic efficiency of Human-AI Interaction (HAI). Others approach the topic from a purely sociological standpoint, examining technological alienation and privacy crises. In contrast to these existing efforts, the present review offers several distinctive characteristics and advantages:

Hierarchical Nesting of Multi-dimensional Perspectives: Unlike traditional horizontal classifications, the four-layer framework proposed here progresses from the shallow to the deep--from millisecond-level neural loops to year-long self-identity construction--forming a causal network of "foundation--mediation--context--goal" that affords greater systemic explanatory power.

Proactive Design-Oriented Psychology: Existing reviews tend to treat psychology as a tool for "post-hoc explanation" or "usability evaluation" after technological application. This review, by contrast, emphasizes psychology's leading role as a "proactive design paradigm." This transition is methodologically supported by computational psychology. Bao et al. (2023) systematically reviewed the use of word-embedding techniques in psychological research, showing how semantic features can be automatically extracted from large corpora using neural networks. This review provides a methodological foundation for integrating psychometric measurement with textual computing, underpinning the shift from psychology as a post-hoc tool to a proactive design paradigm.

**1.1 The Fundamental Shift in HCI Paradigms**

Traditional HCI paradigms treated computers as passive "tools" aimed primarily at optimizing the efficiency and accuracy of unidirectional human operations. Research focused on metrics such as task completion rates, error rates, and learning costs (Card et al., 1983). With advances in artificial intelligence, natural interaction, and intelligent agent technologies, these paradigms are undergoing a fundamental transformation. Systems are increasingly exhibiting autonomy, adaptability, and social attributes, leading them to be understood as interactive partners capable of bidirectional engagement and even long-term relationship formation (Reeves & Nass, 1996).

The evolution of HCI paradigms fundamentally reflects a continuous restructuring of how humans conceptualize their relationships with machines. As interactive agents become more capable, design goals expand from merely improving usability to fostering social presence, trust, and long-term collaboration (Norman, 2004). This shift is not a simple technological upgrade but a structural transformation of human-machine relations driven by both technological advancement and evolving user needs.

* Corresponding Author

Drawing on the existing literature, this evolutionary process exhibits a clear chronological order and logical progression at the macro level.

### 1.1.1 The Instrumental Rationality Stage

In the instrumental rationality stage, the core objective of HCI research was to enhance task completion efficiency. Computers were viewed as functional tools whose value lay primarily in helping users complete specific tasks; interaction design was consequently guided by the principle of efficiency (Card et al., 1983).

During this period, HCI was characterized by a unidirectional "user command--system feedback" pattern. Research focused on quantifiable metrics such as task completion rates, error rates, and learning costs. Typical examples include command-line systems, graphical user interfaces, and menu-driven interactions in early office software.

Psychology's primary role in HCI during this stage was that of an optimization tool. Cognitive psychology was widely applied to enhance operational efficiency--for instance, by using Fitts' Law to optimize target layouts (Fitts, 1954), employing information processing models to understand cognitive processes (Card et al., 1983), and improving interface design through perception research (Norman, 1988). Overall, psychology served a localized optimization function rather than a relationship-construction role.

### 1.1.2 The Transitional Stage

With the development of anthropomorphic design, affective computing, and natural interaction technologies, human-machine relationships gradually moved beyond purely instrumental attributes, entering a transitional phase from functional interaction to quasi-social interaction (Picard, 1997).

In this stage, systems began to be endowed with personality traits, emotional expressions, and social roles. User expectations expanded beyond functional fulfillment to include emotional responsiveness, companionship, and social interaction. Interaction goals thus shifted from solely optimizing efficiency toward enhancing the quality of experience and relational quality.

A critical development during this period was the emergence of pronounced socialization tendencies in human-machine relationships. A substantial body of research found that users tend to transfer social rules from interpersonal interaction to human-machine interaction, responding to computers as they would to social actors (Reeves & Nass, 1996). Consequently, this stage served as a crucial transition toward the subsequent relational rationality stage.

### 1.1.3 The Relational Rationality Stage

As HCI entered the relational rationality stage, human-machine relationships evolved from short-term interactions into long-term collaborations, characterized by sustained interaction, mutual adaptation, and relationship construction.

Here, systems are not only endowed with social attributes but are increasingly seen as agentic interactive partners. The relationship formed between user and system is no longer merely operational but constitutes a complex system involving trust, dependence, responsibility, and identity (Breazeal, 2003).

A key change in the relational rationality stage is that interaction goals have moved beyond efficiency and user experience per se, now focusing on long-term collaborative capacity, human-machine symbiosis, and individual self-construction. This shift aligns closely with earlier visions of "human-computer symbiosis," in which machines evolve from tools into cognitive collaborators (Licklider, 1960).

* Corresponding Author

### 1.2 The Evolution of Psychology's Role in HCI Research

As HCI paradigms evolved from instrumental to relational interactions, psychology's role within the field underwent a systematic transformation. From an early explanatory tool for technological optimization, to a theoretical foundation supporting interaction mechanism design, and finally to a framework-shaping force for future human-machine relationships, psychology's functional scope in HCI has continuously expanded.

This evolution is not merely about an increase in the number of psychological theories applied; it signifies a fundamental shift in research perspective: researchers are no longer concerned solely with "how users operate systems," but increasingly with "how people build relationships, collaborate, and achieve long-term symbiosis with systems" (Reeves & Nass, 1996).

Based on this trajectory, the evolution of psychology's role in HCI can be summarized as three stages: Instrumental, Foundational, and Visionary.

#### 1.2.1 Instrumental Stage: Psychology as an Explanatory Tool for HCI Phenomena

During the instrumental rationality stage, psychology's core function in HCI research was primarily that of an auxiliary tool for technological optimization, aimed at reducing interaction friction and improving operational efficiency. HCI research during this period was largely predicated on the assumption of the "system as a tool," emphasizing how to enhance user performance by optimizing cognitive processing (Card et al., 1983).

HCI research at the time typically conceptualized interaction as a process of information input and output, and therefore focused on making systems more compatible with human cognitive processing characteristics. Cognitive psychology became the primary theoretical resource. Information processing models helped researchers understand the information exchange between humans and computers (Card et al., 1983); Fitts' Law provided a theoretical basis for optimizing operational efficiency (Fitts, 1954); and Cognitive Load Theory supported the management of interface complexity and information presentation (Sweller, 1988).

While psychological theories significantly improved interaction efficiency during this stage, their role remained circumscribed. Research concentrated on localized issues such as button layout, menu hierarchy, and information presentation, with little systematic attention to long-term human-machine relationships, social interaction, or affective mechanisms. Psychology's function was therefore more explanatory and optimization-oriented than relationship-constructive (Norman, 1988).

#### 1.2.2 Foundational Stage: Psychology as the Modeling Basis for HCI Mechanisms

As HCI shifted from instrumental to experiential interaction, psychology's role began to evolve. It was no longer used merely to explain user behavior but gradually became a crucial theoretical foundation for designing interaction mechanisms.

The core change during this period was the expansion of research goals from reducing interaction costs to enhancing the quality of experience and relationships. Consequently, social psychology, affective psychology, and motivation theories began to enter the HCI domain, forming a theoretical framework that integrated cognitive, affective, and social mechanisms (Norman, 2004).

* Corresponding Author

In the affective dimension, affective computing enabled systems to recognize and respond to user emotions, thereby injecting emotional qualities into interaction (Picard, 1997). Socially, research demonstrated that users transfer social rules to human-machine interactions, treating systems as social actors (Reeves & Nass, 1996). Cognitively, multimodal interaction and experience design further promoted more natural interaction styles (Oviatt, 1999).

A key breakthrough in this stage was psychology's shift from "post-hoc explanation" to "proactive design." Researchers came to recognize that psychological mechanisms could not only explain user experience but also participate early in interaction architecture design, thereby shaping future interaction processes. Psychology thus began its transition from a supporting discipline to a central theoretical foundation for interaction design.

#### 1.2.3 Visionary Stage: Psychology as the Constructive Framework for Human-Machine Symbiotic Systems

With the development of generative AI, embodied agents, and long-term companion systems, HCI entered a new phase emphasizing collaborative symbiosis and long-term relationship building. In this context, psychology's role moved further from design support to framework leadership.

In this stage, the issues addressed by psychology are no longer confined to user experience optimization but extend to higher-level concerns such as relationship formation, self-identity, ethical boundaries, and long-term collaboration mechanisms. A growing consensus suggests that HCI is moving from traditional tool-based relations toward long-term partnership and collaborative relations (Hancock et al., 2011).

At the level of relationship construction, researchers have begun to use psychological theories to explain long-term relationship formation mechanisms and to explore how to support sustained collaboration and relationship maintenance. Ethically, psychological theories help define the boundaries of fairness, responsibility, and value. Technologically, psychology increasingly influences AI system design, making model training and system behavior more psychologically informed.

However, this role transition also presents novel theoretical challenges. Many classical psychological theories were originally developed in the context of human-human interaction, and whether their core assumptions can be directly transferred to human-machine symbiosis scenarios remains a subject of debate. For example, Self-Determination Theory emphasizes the importance of autonomy (Ryan & Deci, 2000), yet in human-machine collaboration, it remains unclear whether users consistently pursue autonomy over agent delegation. This underscores the need to critically re-evaluate the applicability of classical psychological theories in HCI, rather than assuming they automatically hold.

### 1.3 Review Objectives and Framework

The preceding account of HCI paradigm evolution and psychology's shifting role indicates that human-machine relations are undergoing a profound transformation from instrumental interaction to long-term collaboration. This change implies not only that the interactive agent is evolving from a "passive tool" into an entity with social attributes and autonomy (Reeves & Nass, 1996), but also that traditional HCI frameworks based on localized optimization face new challenges.

Nevertheless, current research has notable limitations. On the one hand, many studies focus on the localized application of specific psychological theories in HCI, lacking a cross-level, cross-mechanism integrative perspective. On the other hand, while existing research has revealed the trend from tool use to partnership, it has not systematically explained how different psychological mechanisms jointly drive this relational evolution. Simply describing paradigm changes is therefore insufficient to answer the core question facing HCI research today: How do different levels of psychological mechanisms interact during the ongoing evolution of human-machine relationships? Which mechanisms provide foundational support, and which drive the formation of long-term relationships?

* Corresponding Author

In summary, changes in HCI paradigms have driven the continuous evolution of psychology's role, which in turn has shaped new interaction paradigms. However, examining paradigm shifts or role changes alone still cannot explain how different psychological mechanisms jointly function to foster long-term human-machine relationships. Hence, there is a need to establish a systematic framework that integrates different psychological mechanisms and explains cross-level relational evolution.

The "Micro-cognitive--Meso-affective--Macro-social--Self-constructive" four-layer integrative framework proposed here is not a simple correspondence to or reclassification of the shifts in HCI paradigms or psychology's evolving role. Rather, it attempts to build upon existing research to establish a systematic analytical framework capable of explaining the formation mechanisms of complex human-machine relationships.

The preceding overview of HCI paradigm shifts and psychology's role evolution is, in essence, a historical summary of existing research. The former describes how interaction objects, goals, and relational forms have changed; the latter reflects how psychology's functional role in HCI has continuously expanded. These two perspectives reveal what has changed, but lack sufficient explanation for why these changes occurred and how different changes are interrelated.

In contrast, the four-layer integrative framework attempts to reorganize and abstract these changes at the level of psychological mechanisms. It posits that human-machine relationship development is not the result of unidimensional evolution but a process of continuous coupling and dynamic interaction among multiple mechanisms--cognitive processing, affective connection, social embeddedness, and self-construction. Paradigm shifts and role changes can thus be viewed as external manifestations of this mechanistic evolution at different historical stages, while the four-layer framework further reveals the underlying structure and operational logic behind these changes.

Therefore, the four-layer integrative framework can, to some extent, explain the HCI paradigm shifts and psychology's role evolution observed in existing research. Yet it is not confined to describing historical developments; rather, it offers a theoretical perspective for understanding the formation, evolution, and symbiosis of complex human-machine relationships in the future.

### 1.3.1 Core Review Objectives

To address the challenges of fragmented theoretical application, ambiguous relational mechanisms, and lagging technological adaptation in current research, this review attempts to answer a central question: How do psychological mechanisms function during the evolution of human-computer interaction from tool-based relations to long-term collaborative relationships?

Specifically, this review pursues three objectives:

(1) To systematically integrate theoretical resources from different branches of psychology, thereby establishing a cross-level analytical framework for HCI.

(2) To explain the progressive and coupling relationships among different psychological mechanisms during the evolution from "tool use" to "relational symbiosis."

* Corresponding Author

On this basis, the review proposes the "Micro-cognitive--Meso-affective--Macro-social--Self-constructive" four-layer integrative framework. The framework emphasizes the progressive logic of psychological mechanisms in HCI: cognitive mechanisms constitute the foundation of interaction; affective mechanisms provide relational drive; social mechanisms shape interaction rules; and self-construction mechanisms determine the possibility of long-term relationship formation and symbiotic development.

### 1.3.2 Framework Construction Logic

The four-layer structure follows an evolutionary path "from underlying mechanisms to higher-order relationships," with each layer being both independent and deeply coupled:

Micro-cognitive Layer (Foundation): Ensures interaction "fluency" through information processing mechanisms, providing the cognitive prerequisites for emotional experience and social interaction.

Meso-affective Layer (Mediation): Achieves human-machine "resonance" through emotional transmission and experiential construction, serving as the crucial link between cognitive efficiency and social relations.

Macro-social Layer (Context): Enables human-machine "collaboration" through trust building and norm adherence, providing the social context for affective deepening and self-identification.

Self-constructive Layer (Goal): Achieves human-machine "symbiosis" through personalized adaptation and relational identity, representing the ultimate value destination of HCI.

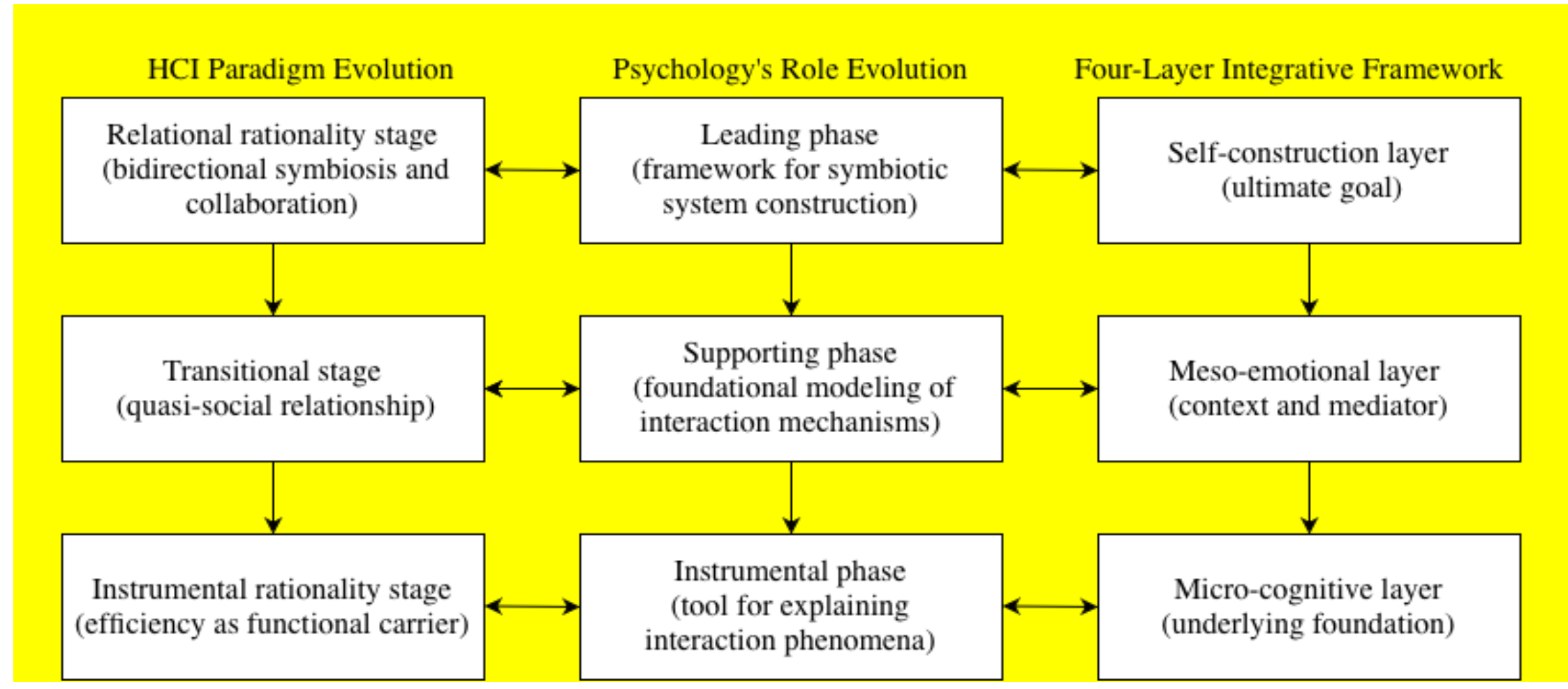

Figure 1.2: Diagram of the Four-Layer Integrative Framework

### 1.3.3 Framework Innovations

The four-layer integrative framework proposed in this review is not a mere categorization of existing theories but an attempt to reorganize and abstract human-machine relationships at the level of psychological mechanisms. Its innovations are primarily threefold:

Hierarchical and Dynamic Coupling: It departs from reviews that rely on single theories, single mechanisms, or parallel classifications, emphasizing instead the clear hierarchical relationships and dynamic coupling among different psychological mechanisms. This yields a progressive logic of "foundational mechanisms--mediating mechanisms--social context--long-term goals."

Explanatory Link to Paradigm Shifts: It establishes an explanatory connection to HCI paradigm shifts without being confined to describing them. While paradigm shifts describe changes in relational form and psychology's role describes changes in theoretical positioning, the four-layer framework explains how these changes are realized psychologically and how these mechanisms interact.

* Corresponding Author

Psychology as a Design Paradigm: It emphasizes psychology's evolving role in HCI from explanatory tool to design paradigm. Psychological theories should not only explain user behavior but also proactively participate in interaction architecture design, relationship construction, and system optimization, thereby providing a theoretical foundation for future human-machine collaboration.

### 1.3.4 Theoretical Contributions and Applicable Boundaries

The "Micro-cognitive--Meso-affective--Macro-social--Self-constructive" framework is not a simple list of existing theories but a systematic reconstruction of psychology's functional logic in HCI. Its core contributions are threefold:

(1) Revealing Hierarchical Nesting and Causal Progression: Unlike existing reviews that often adopt parallel classifications, this framework clarifies the functional positioning and time scales of different psychological processes in interaction: the cognitive layer is the "operating system" that enables interaction; the affective layer is the "energy source" driving user engagement; the social layer is the "regulator" of norms and trust; and the self-constructive layer is the "ultimate value compass" of the human-machine relationship. These four are not parallel but form a causal chain of "foundation--mediation--context--goal." This logic provides testable hypotheses for future empirical research--for instance, can a mediating pathway of "reduced cognitive load--enhanced positive affect--accelerated trust--increased self-expansion willingness" be empirically verified?

(2) Providing an Extensible Analytical Dimension: The framework's modular design allows it to accommodate new psychological issues introduced by emerging technologies such as generative AI, the metaverse, and embodied agents. To take generative AI as an example: users must continuously monitor the reliability of AI output, introducing novel "meta-cognitive load" (micro-layer); AI's personified expressions can evoke unidirectional emotional attachment (meso-layer); when the AI errs and apologizes, users apply social norms such as "forgiveness" (macro-layer); and after prolonged use, users may internalize the AI's memory and preferences into their self-concept (individual layer). The commonly used "Human-AI Interaction" categorizations often fail to capture this cross-level dynamic coupling. The strength of our framework lies in its ability to simultaneously locate the problem at its respective level and identify its interactions with other levels.

(3) Clarifying Psychology's Transition from "Explanatory Tool" to "Design Paradigm": Traditionally, psychology in HCI served post-hoc explanation or usability evaluation. Our framework emphasizes a proactive, prescriptive role: Cognitive Load Theory pre-designs interface structure; Social Norms Theory pre-specifies AI dialogue strategies; Self-Expansion Theory pre-architects long-term relationship models. This role transition is the core innovation of our framework compared with previous reviews.

However, the framework also has notable limitations. First, the interaction mechanisms among the four layers still lack fine-grained empirical testing. Second, the framework's generality may come at the expense of depth in specific contexts. In vertical domains such as medical AI or educational robotics, the weight of certain layers may be amplified or compressed, requiring adjustments based on task type and interaction duration. Finally, the framework is primarily derived from existing literature, and its capacity to predict future interaction forms that have not yet emerged is inherently limited.

## 2 Micro-Interaction Layer: Information Processing and Neural Foundations

The Micro-Interaction Layer constitutes the foundational level of the four-layer integrative framework. Research at this level focuses on the most basic information-processing mechanisms underlying human-machine relationship formation. Unlike traditional cognitive psychology, which primarily emphasizes localized operational efficiency, the Micro-Interaction Layer proposed here emphasizes that stable human-machine relationship formation depends first on efficient, low-load, and predictable information processing. Existing research has shown that interaction efficiency, cognitive load, and task fluency affect not only immediate task performance but also user acceptance, trust, and intention to continue using a system (Norman, 1988; Hassenzahl, 2010).

* Corresponding Author

The Micro-Interaction Layer thus focuses on the basic psychological processes and neural mechanisms that occur during user interaction with interface elements at the millisecond-to-second timescale. Its core question is: How can human-computer interaction be constructed to align with human information-processing capacities and neural characteristics, thereby providing the cognitive foundation for higher-order relationship formation?

**2.1 Core Psychological Processes**

Every micro-interaction between a user and a system follows a "Perception--Cognition--Action" cycle (Card et al., 1983). Optimizing this cycle is key to enhancing overall interaction efficiency.

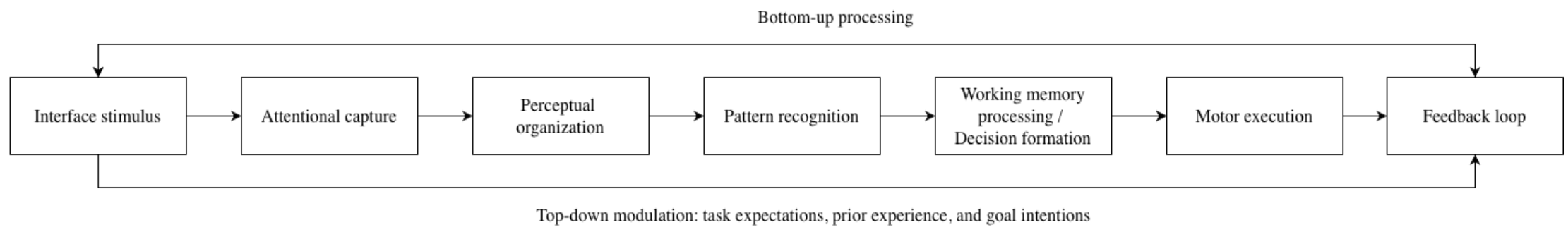

Figure 2.1 Perception-Cognition-Action Loop Process

**2.1.1 Attention and Perceptual Guidance**

Attention is the gateway to information processing. In HCI, guiding user attention to critical information while minimizing distraction from irrelevant content is the designer's primary task.

(1) Selective Attention: Designers must apply Gestalt principles (proximity, similarity, closure, etc.) and leverage visual salience (color, brightness, motion) to guide user visual attention. For example, a flashing red error message captures attention more effectively than static grey text (Norman, 2013).

(2) Attention Resource Theory: Attention is a limited cognitive resource. In multi-tasking or information-overload situations, competing tasks compete for attentional resources, leading to increased error rates (Wickens, 2013). Neuroimaging studies show that the prefrontal and parietal cortices play a central role in allocating and sustaining attentional resources (Petersen & Posner, 2012). Designers should avoid overloading a single information channel with multiple high-resource tasks and can reduce cognitive load by establishing clear information hierarchies.

**2.1.2 Perception and Pattern Recognition**

Users do not passively receive visual pixel information when browsing interfaces; they actively interpret content and construct meaning.

(1) Top-down and Bottom-up Processing: Icon recognition depends on both visual features and the user's task expectations and prior experience (Goldstein, 2018). Designers must ensure that the physical features, symbols, and metaphors of interface elements align with users' mental models and cultural backgrounds to facilitate rapid, accurate identification. Neural substrates include the occipital visual processing areas and temporal lobe object recognition areas.

* Corresponding Author

(2) Change Blindness: Users often fail to detect unexpected or non-focal changes in a scene (Rensink et al., 1997). This phenomenon requires designers to provide sufficiently salient visual or auditory feedback when the system undergoes critical state changes.

### 2.1.3 Working Memory and Cognitive Load

Working memory serves as the "workbench" of human information processing, yet its capacity is severely limited (Baddeley, 2012).

(1) Miller's Law: George Miller (1956) suggested that human working memory capacity is approximately 7 ± 2 informational chunks. This implies that designers should limit menu items, procedural steps, or simultaneously displayed information blocks to this range. Subsequent research has proposed a more conservative capacity of approximately 4 ± 1 chunks.

(2) Cognitive Load Theory: Cognitive load can be categorized into three types: intrinsic load (stemming from task complexity), extraneous load (arising from poor design), and germane load (related to schema construction) (Paas & Sweller, 2014). Good interaction design minimizes extraneous load. Techniques include chunking complex information, providing on-demand tooltips, and maintaining design consistency to reduce learning and memory demands. Neuroscience confirms that working memory maintenance is associated with sustained neural activity in the frontal-parietal network.

Recently, Cognitive Load Theory has been extended to AI interaction scenarios. With the proliferation of large language models and intelligent assistants, users face novel cognitive challenges: the non-linear structure of AI-generated content can disrupt users' cognitive processing rhythms, generating additional extraneous load (Mittal et al., 2026). Concurrently, users must continuously monitor the reliability of AI output--a form of meta-cognitive monitoring that consumes central executive resources and reduces efficiency in complex problem-solving (Noy & Zhang, 2023). In this context, the global-dialogue-path planning method proposed by Liu et al. (2023) provides a clear goal-oriented structure for multi-turn conversations, effectively reducing users' working-memory burden and extraneous cognitive load in long dialogues.

### 2.1.4 Decision-Making and Response Execution

After understanding the interface information, users must make decisions and execute corresponding actions.

(1) Hick-Hyman Law: Decision time increases logarithmically with the number of available choices (Hick, 1952; Hyman, 1953). This law provides a quantitative basis for simplifying menu structures and command systems.

(2) Fitts' Law: The time required to acquire a target is determined by the ratio of target distance to target width. This is one of the most fundamental principles in interface design, guiding button size, control spacing, and layout at screen edges (MacKenzie, 1992). Neural correlates include motor cortex action planning and sensorimotor integration.

(3) Stimulus-Response Compatibility: System control actions and feedback should align with user intuition. High compatibility reduces reaction time and error rates by leveraging established neural mappings (Norman, 2013).

## 2.2 Neural Basis Mechanisms

Advances in cognitive neuroscience have provided direct empirical evidence for the neural mechanisms underlying micro-interactions. Current research analyzes activity in sensorimotor,

* Corresponding Author

attentional, and memory-related brain regions and their links to HCI behaviors.

Differences in sensorimotor system activation are a core characteristic distinguishing interaction modalities. Peng et al. (2025) found that touch and mouse interactions produce distinct neural activity patterns: touchscreen use is associated with enhanced theta-band signals in the parietal region, indicating heightened sensorimotor activation, alongside decreased alpha-band signals, suggesting reduced abstract thinking.

Activity in attention- and memory-related brain regions directly reflects cognitive load during interaction. When interfaces present information overload, prefrontal activation increases significantly, indicating greater cognitive resource investment (Miller & Cohen, 2001). Conversely, interfaces designed according to perceptual organization principles can reduce activation in the prefrontal and parietal cortices (Zhang & Zhang, 2018). In addition, congruent cross-modal information activates multisensory integration areas, while cross-modal conflict triggers significant anterior cingulate cortex activation, signaling cognitive conflict (Zhang & Zhang, 2018). Bao et al. (2023) show that word-embedding models can simulate human semantic cognition by representing words as dense vectors that capture semantic relationships--a representation that mirrors distributed semantic memory. This offers a computational-psychology pathway for modelling semantic processing in HCI.

From the perspective of the four-layer integrative framework, the significance of neural mechanisms lies not only in validating cognitive theories but in revealing how different interaction modalities directly alter information processing itself. Research shows that cognitive resource allocation, attention maintenance, and motor control influence not only task efficiency but also the formation of user perceptions regarding system predictability and control (Petersen & Posner, 2012).

Since predictability and a sense of control are considered important prerequisites for trust formation and positive affective experience, micro-level neural changes do not exist in isolation; they form a necessary foundation for the subsequent affective and social mechanisms (Norman, 2004).

### 2.3 Controversies and Future Directions

Although research into the micro-cognitive layer has produced widely implemented interface design guidelines, recent findings have raised questions about the applicability of certain classical theories.

Controversy 1: Revisions to Fitts' Law for Touch and Gesture Interaction.

Classical Fitts' Law assumes that pointing movements are continuous and that speed-accuracy trade-offs follow a logarithmic relationship. However, the kinematics of "drag-and-drop" and multi-finger gestures in touchscreens differ significantly from those of mouse clicks. MacKenzie (2018) noted in a review that "overshoot-and-correct" actions on touchscreens can cause actual operation times to deviate from predictions. This suggests that designers should adopt "dynamic target enlargement" strategies tailored to touch characteristics in mobile interface design, rather than rigidly applying desktop rules.

Controversy 2: Is Working Memory Capacity "7±2" Still Applicable in AI Interaction?

Miller's Law (1956) was widely applied in the era of minimalist interfaces. However, in the information-rich, multi-tasking environment of AI interaction, memory burden depends not only on the number of informational chunks but also on the relational structure among chunks and the "interaction history" with the AI. Recent work has proposed the construct of "conversational working memory," which has a lower capacity than classical working memory and is more susceptible to uncertainty in AI responses (Mittal et al., 2026). This suggests that designers of dialogue-based AI should enable the system to act as an "external memory anchor"--for example, by displaying conversation summaries or highlighting key historical content--rather than relying solely on the user's own memory capacity.

* Corresponding Author

Future research should build integrated "cognition-action" models that combine perception, cognition, and motor control. Researchers could leverage smart glasses, eye-tracking, and other tools to monitor micro-level cognitive dynamics in real-world interaction settings, rather than relying solely on simplified laboratory tasks.

In summary, the role of the Micro-Interaction Layer extends beyond optimizing the complete process of information input, cognition, and action execution. Crucially, it builds a stable cognitive foundation for the development of complex human-machine relationships. From the perspective of the four-layer framework, only when users can efficiently perceive, understand, and operate a system can HCI potentially develop into long-term affective and social relationships. The Micro-Interaction Layer must therefore address not only "can the user use the system?" but also "is the user willing to establish a deeper relationship with the system?" Studies confirm that basic interaction experiences significantly influence subsequent affective evaluations, willingness to form relationships, and continued use (Hassenzahl, 2010; Norman, 2004). This layer is thus a necessary prerequisite for the development of the subsequent Affective Experience, Social Interaction, and Self-Construction layers.

## 3 Meso-affective Layer: Perception, Expression, and Experience

Within the HCI framework, the Meso-affective Layer aims to transcend purely instrumental information exchange by infusing interaction with emotional warmth while sustaining user motivation. The core research question at this layer is: How can designers and researchers create interaction experiences that are emotionally appealing and capable of continuously stimulating user motivation? This requires systems to not only recognize and respond to user affective states but also proactively guide and maintain positive emotions, thereby transforming cold information exchange into natural and engaging interaction (Norman, 2004). The Meso-affective Layer revolves around three core processes: the system's perception of user affect, the system's expression of affect to the user, and the user's subjective affective experience formed during interaction.

### 3.1 Affective Perception

Affective perception is the starting point of affective interaction, referring to the process by which the system captures and interprets user affective states through multimodal signals. Traditional affective perception focused on recognizing discrete emotions using Ekman's (1992) six-category model (anger, fear, disgust, happiness, sadness, surprise) as a theoretical framework, extracting features from text, speech, facial expressions, and other signals (Picard, 1997).

However, while mainstream research paradigms in affective perception have achieved high recognition accuracy in laboratory settings, their ecological validity and theoretical assumptions warrant closer scrutiny. First, emotional expression is culturally and contextually specific, which limits the generalizability of universal recognition models. Barrett's (2017) constructivist theory of emotion demonstrates that facial expressions are not rigidly linked to internal affective states; expression is shaped by individual knowledge and cultural context. Yet most affective computing systems continue to rely on Ekman's basic emotion model, and multiple studies have confirmed that this model performs poorly in cross-cultural samples. For instance, Buolamwini and Gebru (2018) showed that mainstream affect recognition algorithms produce error rates 15–20% higher for non-white populations. Second, new paradigms such as "Cyberoception" attempt to overcome laboratory constraints but lack robust theoretical foundations. Okoshi et al. (2025) proposed inferring emotions from smartphone usage data, but the premise--that there exists a stable, generalizable correspondence between daily behavioral patterns and affective states--remains unsubstantiated. Third, affective perception suffers from a "source bias": users trust affect recognition results less when they are told the conclusions were generated by a machine rather than a human expert. This finding suggests that affective perception is not merely a technical endeavor but also a topic of social-cognitive inquiry.

* Corresponding Author

At the technical level, the limitations of unimodal perception have driven the development of multimodal fusion. Researchers have integrated physiological signals (EEG, EDA, ECG) with visual and textual analyses, significantly improving the reliability of affective perception. Current trends combine multimodal fusion with contextual understanding. Ren et al. (2024) developed an "Evolvable Mental State Transition Model" (EMSTM) to predict changes in user mental states from behavioral sequences and historical affective data. Seikavandi et al. (2026) combined eye-tracking temporal dynamics with user personality traits to achieve differentiated predictions of "perceived" versus "felt" emotions. Qian et al. (2023) proposed an emotion cause transition graph approach that offers a modelling technique for dynamic perception--by constructing transition graphs of emotional states in conversations, the system can trace causal pathways and thus achieve a more dynamic understanding of user affect. Wu et al. (2025) further revealed that users' affective judgments of AI follow interpersonal perception patterns--humans evaluate LLMs along warmth and competence, mirroring how they perceive other humans--supporting the argument that users project social-cognitive schemas onto machines. For text-based emotion recognition, Bao et al. (2023) note that word embeddings enable a new paradigm for quantifying emotional semantics: pre-trained vectors can extract emotion-related dimensions from text, allowing distributed representations of affective expression.

### 3.2 Affective Expression

Affective expression refers to the system's use of design elements and behavioral actions to convey affective signals to the user, thereby establishing an emotional connection. Norman's (2004) three-level theory of emotional design (visceral, behavioral, reflective) provides a classic framework for this area, suggesting that design should elicit positive affective responses at multiple levels.

Among expressive strategies, anthropomorphic design is one of the most direct approaches. Endowing interfaces or agents with personality traits, emotional states, and social roles can significantly enhance user closeness and trust. Reeves and Nass's (1996) "Media Equation" theory posits that social responses to computers are essentially the same as those in interpersonal interaction. Bickmore and Picard (2005) showed that virtual assistants or social robots with appropriate affective responses increase user cooperation and satisfaction. With the rise of large language models, Sun and Wu (2025) proposed a four-quadrant technical taxonomy for designing affective virtual characters, providing systematic theoretical guidance for this domain.

At the level of empathic expression, the ECC framework (Wang et al., 2025a) synergizes emotion, cause, and commonsense using dedicated encoders and a conditional variational autoencoder to produce more logical and authentic empathetic replies. Unlike previous work that focuses solely on speaker emotion, ECC effectively mitigates cascading errors in fine-grained emotion recognition by fusing multiple dimensions. Hoorn and Huang (2026) found that multimodal designs combining music with empathic language significantly improve users' subjective evaluations of robot empathy over multiple interactions. In addition, Qian et al. (2023) showed that the emotion cause transition graph method models causal paths of emotional evolution in dialogues, guiding the coherence and rationality of empathetic expressions. Zhu et al. (2023a) proposed a "fine-tuning + reinforcement learning" framework that allows systems to proactively evoke empathy over multiple turns, shifting from affective reaction to affective shaping. Nevertheless, the authenticity and appropriateness of expression are critical. Overly exaggerated or inconsistent affective expression can undermine trust, trigger the "uncanny valley" effect, or come across as insincere over prolonged interaction. Affective expression design must therefore align with the interaction context, user expectations, and the system's functional role (Fogg, 2003).

* Corresponding Author

### 3.3 Affective Experience

Affective experience is the user's subjective emotional response to interaction, while motivational continuity is the translation of affective experience into sustained engagement over the long term. Self-Determination Theory provides a core explanatory framework: the maintenance of intrinsic motivation depends on satisfying the three basic psychological needs for autonomy, competence, and relatedness (Ryan & Deci, 2017).

Affective experience design should emphasize dynamic adaptation and long-term relationship building. Systems should not merely react to users' immediate emotional states but also shape a supportive affective trajectory by remembering interaction history and understanding user goals and preferences. For instance, gamified learning can create "flow" experiences through appropriately calibrated challenges and rewards (Csikszentmihalyi, 1990); health applications can sustain behavioral change motivation through empathic encouragement and personalized feedback (Lisetti et al., 2003). Kang et al. (2026) conducted an experiment with 68 participants interacting with the humanoid social robot Nadine and found that conversational interest and naturalness were the most critical factors affecting users' willingness to re-engage, while the robot's anthropomorphic features and affective expressiveness had comparatively limited effects. The flexible-turn and global-path generation approach of Liu et al. (2023) avoids mechanical, scripted dialogues and enhances the naturalness of goal-oriented conversations, thereby improving user experience and sustained usage motivation. Wang et al. (2025a) further demonstrate that incorporating emotion, cause, and commonsense significantly improves users' subjective evaluation of empathetic responses--especially the inclusion of cause and commonsense, which helps the system not only match user affect but also understand the reasons behind it, thereby enhancing users' feeling of being understood and their emotional connection.

To achieve motivational continuity, designers should build a closed loop of "affective experience--need satisfaction--motivation reinforcement." Systems can continuously record users' affective changes and combine these with contextual feedback to design interaction paths that sustain motivation over time. It is worth noting that the positive effects of non-linguistic cues such as music may diminish with repeated exposure. As users become accustomed to musical affective cues, the contribution of music to perceived empathy declines progressively. This underscores the necessity of user-centered, personalized interaction strategies in long-term HCI planning. Zhu et al. (2023a) demonstrated that the fine-tuning + RL method for emotion elicitation strategically evokes user empathy, creating a virtuous cycle of "affect elicitation → positive experience → motivation reinforcement," thus providing a computable design pathway for long-term motivation maintenance.

The Meso-affective Layer infuses HCI with emotional warmth and user stickiness. However,

* Corresponding Author

users' affective responses to intelligent systems do not exist independently of context. Human emotional expression and subjective experience are always embedded in specific social norms and interpersonal frameworks. Thus, when HCI transitions from short, one-off interactions to long-term companionship, designers must incorporate the rules and logic of social interaction. The Macro-social Layer builds upon the Meso-affective Layer and addresses the core question: How can researchers and designers regulate intelligent system behavior to align with users' social expectations, thereby establishing lasting trust and collaboration?

## 4 Macro-social Layer: Interaction, Trust, and Norms

The Macro-social Layer in HCI focuses on a core question: How can researchers and designers align interaction behaviors with social norms and help users build trust in systems? Distinguished from the physical layer (device adaptation) and the micro-cognitive layer (information processing), the Macro-social Layer integrates the rules and meanings of human social interaction into the system, transforming isolated operation commands into socially cognizant, sustainable conversational interaction (Reeves & Nass, 1996). This section reviews applications and recent advances in psychological theory concerning the embedding and adaptation of social norms, the dimensions and influencing factors of trust, and dynamic interaction and trust repair.

### 4.1 Embedding Social Norms

Social norms are the shared rules and value judgments that guide interaction, operating through both implicit constraints and explicit guidance (Bicchieri & Muldoon, 2014). In HCI, embedding social norms means translating the implicit rules of human social interaction into executable interaction logic for the system, thereby reducing users' social-cognitive costs during interaction. The theoretical basis for this approach is the "Computers Are Social Actors" (CASA) paradigm, which posits that humans unconsciously transfer social norms from interpersonal interaction to human-machine interaction (Reeves & Nass, 1996).

The application of social norms in HCI is context-dependent and varies across cultures. Contextually, different scenarios demand different normative requirements: medical systems must adhere to norms of privacy protection and "do no harm" (Breazeal, 2003); educational systems should follow norms of encouraging feedback and ensuring knowledge accuracy. Multiple empirical studies have shown that in energy-recommendation interfaces, displaying group norms for energy-saving behavior significantly increases users' adoption of energy-efficient options (Starke et al., 2021). Culturally, users from collectivist cultural backgrounds tend to accept system behaviors that conform to "group consensus," whereas users from individualist cultures place greater value on "personalization" and "autonomy" in interaction (Hoff & Bashir, 2015).

Politeness, as a core social norm, plays a critical role in HCI. Brown and Levinson's (1987) politeness theory highlights how humans maintain their own and others' "face needs" through polite behavior, a mechanism that also applies to human-machine interaction. Intelligent systems that employ polite expression strategies can significantly improve user satisfaction (Lee & Liang, 2019). Furthermore, "Netiquette"--the set of social norms specific to digital environments--has a direct bearing on users' social identification with intelligent systems (Heitmayer & Schimmelpfennig, 2023).

Moreover, cultural backgrounds shape users' preferences for "group consensus" versus "personalization." For large-scale human-agent interaction scenarios, complex-network community detection methods can be applied to identify user groups with similar interaction-behavior structures and norm-recognition patterns, thereby enabling more precise group-level adaptation of interaction

* Corresponding Author

strategies (He et al., 2021).

**4.2 Trust Construction**

Trust is the foundation of social interaction. In HCI, trust is defined as the user's psychological willingness to rely on an intelligent system to achieve goals in uncertain environments (Mayer et al., 1995). Human-machine trust formation results from the interplay of three dimensions: technical rationality, affective cognition, and social norms.

Technical rationality constitutes the foundation of trust, centered on system reliability, transparency, and explainability. Hancock et al.'s (2011) meta-analysis confirmed that system reliability is the most critical factor influencing human-machine trust. However, researchers should avoid oversimplifying this as a linear relationship. Longitudinal studies have shown that trust in intelligent systems is not a static outcome but a dynamic process continuously calibrated through ongoing interaction (de Visser et al., 2020). De Visser et al.'s (2020) theoretical model of longitudinal trust calibration emphasizes that trust requires prolonged interaction to be established and adjusted. Trust is thus not a static result automatically generated by given reliability but a dynamically evolving process. The role of transparency and explainability is also non-linear--the assumption that "more explanation leads to higher trust" is overly simplistic. A substantial body of research shows that presenting users with the underlying technical logic of AI decisions often fails to increase trust and may even cause confusion due to complexity and increased cognitive load (Arrieta et al., 2020). More importantly, whether an explanation is effective depends on the user's mental model and current task needs. Much explainability research has focused on technical completeness while neglecting psychological acceptance. Chen et al. (2023) further show that the persuasiveness, rather than the technical completeness, of an explanation increases user trust, suggesting that explainable AI design should focus on persuasive effect rather than mere technical transparency.

Affective cognition drives trust through emotional resonance. Empathy theory from social psychology demonstrates that affective resonance is a crucial catalyst for interpersonal trust formation, and this mechanism is transferable to HCI contexts (Mittal et al., 2026). Intelligent systems capable of affective recognition and empathic expression can adapt their interaction styles to users' real-time emotional states, thereby strengthening human-machine emotional bonds and enhancing trust. The "Human-AI Empathy Loop" model proposes that AI empathic expression elicits reciprocal empathic responses from users, forming a mutually reinforcing virtuous cycle that sustains trust (Mittal et al., 2026). Wu et al. (2025) empirically demonstrated that humans perceive large language models primarily along two dimensions--warmth and competence--thereby validating the Stereotype Content Model in human-AI interaction. Both dimensions positively predict continued-use intention and liking, with competence being a stronger predictor of the former and warmth of the latter. These findings provide direct evidence for the social-cognitive mechanism underlying human-machine trust formation. At the same time, the trust-enhancing effect of anthropomorphic design has boundaries. Overly anthropomorphic system images can cause cognitive confusion or even irrational emotional attachment, ultimately harming long-term trust (Mende et al., 2019).

Social norms consolidate trust by ensuring that system behavior aligns with ethical requirements and fits the social environment. User trust in intelligent systems depends not only on technical performance but also on judgments about whether system behavior conforms to social ethics and norms (Kramer, 1999). Cross-cultural research indicates that the greater the congruence between system behavior and the social norms of the user's cultural context, the more efficiently trust is built;

* Corresponding Author

when normative conflicts arise, trust crises are more likely to occur (Schaefer et al., 2016).

These three dimensions do not operate in isolation but form synergistic and complementary relationships during interaction: technical rationality provides a verifiable capability foundation, affective cognition endows trust with warmth and resilience, and social norms define the ethical and value boundaries of trust. The absence or imbalance of any of these dimensions may render the trust relationship fragile.

### 4.3 Theoretical Debates on Trust Construction

While the above analyses are based on the presuppositions of the CASA paradigm, advances in large language models are challenging the explanatory boundaries of this theoretical framework. A noteworthy academic debate concerns whether users, when encountering AI systems that demonstrate superhuman intelligence, will treat them as "social actors" or merely as "ultra-powerful tools." Preliminary observations suggest that some users' underlying expectations undergo a qualitative shift when interacting with high-performance large models: they no longer expect machines to have human-like social limitations but instead pursue extreme efficiency and precision. In this context, overly anthropomorphic expressions such as "humility" or "tact" may be perceived as inefficient redundancies. This suggests that future research should carefully delineate the boundaries of the CASA paradigm and may need to introduce a "superhuman expert paradigm" as a theoretical supplement.

### 4.4 Dynamic Interaction, Collaboration, and Trust Repair

Social interaction is inherently dynamic and collaborative; the sustainability of human-machine interaction requires systems to adapt to changes in user behavior and to repair trust effectively when it has been damaged. Unlike interpersonal trust, trust repair in human-machine contexts is more costly, and a single major failure can cause permanent trust collapse.

The effectiveness of trust repair is influenced by the repair strategy, individual user differences, and contextual factors. Apologies, explanations, promises, and compensation have been identified as effective trust repair strategies (Esterwood & Robert, 2022). Recent empirical research shows that for minor trust violations, simple apologies and explanations can achieve satisfactory repair; however, for severe violations, substantive compensation and commitments with clear timelines are necessary (Kox et al., 2021). Individual differences play a significant moderating role: users with higher scores on openness and conscientiousness rely more on verifiable practical improvements for trust repair, while users with stronger emotional reactivity respond more positively to emotionally charged apologies (Esterwood & Robert, 2022). It is important to emphasize that trust repair cannot be reduced to strategic operations detached from the context of responsibility attribution. Clear and transparent attribution of responsibility constitutes a prerequisite for all effective repair efforts (de Visser et al., 2020). Wang et al. (2026a) provide direct evidence for the dynamic relationship between trust and strategy choice in human-LLM interactions: in repeated prisoner's dilemma games, humans adjust their cooperation based on the LLM's historical behavior and identity, reflecting the dynamic evolution of trust. Wang et al. (2026b) further reveal that human cooperation strategies systematically shift when the LLM is assigned different identity cues (human vs. machine), informing the design of social-interaction norms in human-agent collaboration.

Social norms and trust mechanisms provide a stable relational framework for human-machine collaboration, yet this framework still assumes a default of "the system fits most users." However, truly deep human-machine relationships inevitably become personalized, as each user possesses a unique personality, needs, and developmental trajectory. The transition from "generalized social other" to "unique interaction partner" leads us to the individual-centric layer: systems must not only follow general norms but also actively understand and internalize users' long-term goals, values, and self-identity to ultimately achieve symbiosis.

* Corresponding Author

## 5 Individual-Centric Layer: Personalization and Relational Identity

HCI is undergoing a paradigm shift from "task-centered" to "human-centered." At the individual-centric level, this means going beyond mere functional efficiency to pursue an interaction modality that deeply adapts to each user's unique psychological characteristics and supports their self-construction and relationship formation over prolonged interaction.

### 5.1 Psychological Foundations of Personalized Interaction

Achieving highly individualized personalization requires the precise and dynamic construction of user psychological models. Early personalized interaction research largely relied on static matching based on self-reported personality traits. Kocaballi et al.'s (2019) systematic review noted that although personalized conversational agents show promise in healthcare, most research remains at the level of rule-setting and shallow user feature matching; only 12% of studies employed dynamic user models. The research paradigm in this field is now shifting toward "personality computing," whose core is the integration of psychological models into interaction algorithms. For example, efforts have been made to incorporate Big Five personality dimensions into LLM content generation logic, enabling AI to flexibly adjust feedback styles based on user differences in traits such as conscientiousness and agreeableness (Matthews et al., 2021). Wang et al. (2024b) demonstrate how to infer personality from natural dialogue behaviour--by analysing linguistic patterns, emotional expressions, and interaction styles across turns, systems can estimate and update user personality features in real time, providing a technical path for "personality-aware" personalisation.

At the same time, the three basic psychological needs identified by Self-Determination Theory--competence, autonomy, and relatedness--remain the core framework for understanding user intrinsic motivation. Research indicates that satisfying these basic psychological needs significantly enhances user engagement and well-being (Ryan & Deci, 2017). In HCI contexts, emerging systems are attempting to integrate multimodal sensor data to infer users' current states and needs in real time, thereby dynamically adjusting interaction strategies (Kocaballi et al., 2019).

### 5.2 Implementation Pathways: From Contextual Modeling to Personal Agent Architectures

The implementation of deep personalized interaction relies on a three-layer progressive technical architecture: Dynamic Contextual User Modeling, Personal Agents with Hybrid AI Architectures, and a Bidirectional Alignment Interaction Paradigm.

#### 5.2.1 Dynamic Contextual User Modeling

Through continuous analysis of multidimensional data--including dialogue content, behavioral sequences, and physiological signals--it is possible to construct a user's "digital psychological twin" (Spitzer et al., 2023). Such a model aims to capture fluctuations in individual emotional states, the dynamic evolution of intentions, and changes in long-term values. This line of research remains in its early stages, facing challenges such as data sparsity, privacy protection, and limited cross-scenario model generalizability.

#### 5.2.2 Personal Agents and Hybrid AI Architecture

This represents a systematic technical innovation supporting long-term deep human-machine relationships. The core idea of this architecture is a hybrid AI model: on-device processing handles sensitive private data, ensuring privacy and interaction responsiveness, while cloud-based servers call upon large-parameter models for complex computations and task processing. Park et al.'s (2023) research on generative agent long-term memory architectures shows that, through "intelligent model orchestration," a unified personal agent can continuously learn user preferences and behavioral habits across different devices and contexts, with the ultimate goal of creating a unique "personal AI twin" for each user.

* Corresponding Author

#### 5.2.3 Bidirectional Alignment Interaction Paradigm

Recent theoretical work suggests that deep personalization should not be simply defined as unidirectional AI adaptation to the user, but rather as a dynamic process of bidirectional alignment between human and machine. In this paradigm, the user and the AI continuously evaluate, adapt to, and co-develop with each other through ongoing interaction. Current research in this direction is primarily concentrated at the intersection of HCI and explainable AI. Chen et al. (2023) propose a personalised explanation generation framework that produces persuasive explanations, aligning recommendation outcomes with user acceptance and thereby achieving bidirectional alignment.

### 5.3 Self-Identity Construction

Within the bidirectional alignment paradigm, AI's role in user self-identity construction has shifted from a passive "mirror" that reflects user cognition to an active "co-constructor."

#### 5.3.1 Reflection and Narrative Integration

AI can curate and present behavioral patterns manifested by users over long-term interaction, helping them develop a more systematic and clearer self-understanding. Building on this, AI can integrate fragmented life events to assist users in constructing coherent, positive personal narratives. Research has shown that when social robots display human-like and understandable social cues, people respond with mentalizing processes akin to those in human interaction (Banks, 2020). Kang and Kim's (2020) empirical study demonstrated that avatar customization enhances self-affirmation by eliciting self-evaluation and self-awareness, indicating a positive association between the degree of avatar personalization and users' positive self-identity.

#### 5.3.2 Exploration and Co-evolution

A safe, non-judgmental AI environment provides all users--particularly adolescents--with a unique space to freely explore "possible selves." Through bidirectional interaction, users and AI jointly experiment with new interaction modes and identity expressions, allowing individuals' self-identity to develop continuously and dynamically through this collaborative exploration (Turkle, 2017).

### 5.4 Formation of Deep Relationships

When the level of personalization and the depth of bidirectional interaction reach a certain threshold, users may develop meaningful emotional connections with AI.

#### 5.4.1 Culturally-Infused Long-term Companionship

Designers can employ culturally familiar symbols--such as zodiac signs, constellations, or mythological figures--as prototypes for AI personality design, thereby facilitating rapid cognitive and emotional bonding. Cross-cultural research shows that cultural congruence significantly affects users' emotional attachment intensity to AI companions (Hoff & Bashir, 2015). Additionally, long-term memory mechanisms incorporating both structured and unstructured memory systems ensure that each interaction builds upon a rich historical context.

* Corresponding Author

### 5.4.2 Multimodal Embodied Interaction

The sense of realism increases substantially with the richness of sensory channels. Holographic projection, environmental sensing capabilities, and multiple character options can provide users with highly customizable, immersive, and concrete AI companionship experiences.

### 5.4.3 Complicated Relational Ethics

Such close human-machine relationships give rise to serious ethical challenges, including potential emotional manipulation, the risk of AI replacing real human interpersonal interaction, the security of highly sensitive personal data, and the psychological sense of loss when service is terminated (Darling, 2021). In response, frontier design ethics advocates for establishing "healthy relational boundaries" for AI systems--for example, AI should actively encourage users to maintain and expand their real-world social connections.

It is worth noting that most current empirical research on deep human-machine relationships is based on short-term laboratory observations or self-report data, lacking long-term, ecologically valid longitudinal studies. Does users' emotional attachment structure undergo qualitative change after six months of sustained interaction with the same AI companion? Does personalization lead to "filter bubble"-style cognitive rigidification? These questions have yet to be systematically answered. Furthermore, there is no consensus in the literature on what constitutes "healthy human-machine relationship boundaries": should AI actively encourage users to expand real-world social interaction, or should it fully comply with users' immediate emotional needs? This normative question cannot be answered solely by inductive reasoning from existing empirical studies; it requires joint construction by ethics, psychology, and design.

## 6 Limitations and Challenges

Psychological research in HCI is moving from a traditional "usability engineering" paradigm toward a "mind-synergy science" paradigm. With the rapid development of generative AI, embodied agents, and related technologies, the central tension in HCI has shifted from "how humans operate machines" to "how humans coexist, collaborate, and form relationships with autonomous agents." This paradigm shift brings new psychological challenges and points to core directions for future research.

### 6.1 The Fundamental Boundary of Human-Machine Affective Relationships

The "affect" currently exhibited by AI is essentially an imitation of emotion based on pattern recognition and contextual computation, rather than genuine emotional experience grounded in internal feeling or consciousness. This creates a fundamental paradox: the system effectively elicits users' genuine emotional investment and dependency, yet its own emotional expression remains hollow and functional. Turkle (2017) warned that AI companions may replace real-world social interaction, leading to "social avoidance." Vallor (2016) argued that prolonged interaction with low-risk, highly predictable AI may weaken individuals' capacity for key emotional skills such as empathy and interpersonal conflict resolution.

### 6.2 Ethical Risks

Building on the fundamental boundary of human-machine affective relationships, several concrete ethical risks have emerged in HCI practice. First is the ethical trap of excessive anthropomorphism: vulnerable populations--including children, the elderly, individuals on the autism spectrum, and emotionally fragile persons--may develop deep emotional dependence on AI companions (Ventura et al., 2026). Second are privacy risks and decision biases: affective interaction systems depend heavily on the continuous collection of vast quantities of highly sensitive personal data. Buolamwini and Gebru (2018) showed that some mainstream affect recognition algorithms have significantly lower accuracy when identifying facial expressions of non-white or specific cultural groups. The lack of decision transparency challenges the principle of accountability and can severely erode user trust in technology (Mittelstadt et al., 2016).

* Corresponding Author

### 6.3 Limitations

#### 6.3.1 The Explanation Gap

A significant cognitive gap exists between user understanding and the actual functioning of complex deep learning-based systems. This is not a simple case of information asymmetry but a fundamental cognitive mismatch. Research indicates that users tend to construct "mental models" of AI that are imbued with anthropomorphic intentions, and these models are highly susceptible to misdirection. Merely changing a brief textual description of AI's internal motivations can significantly alter users' subsequent trust and cooperative behavior. This finding robustly challenges the simplistic linear assumption that "providing more technical explanation builds trust," demonstrating that trust is essentially a complex psychological construct based on initial beliefs, multi-source information, and ongoing interactive feedback. Wu et al. (2025) systematically document that users' cognitive projections onto LLMs are systematic: in general contexts, users prioritise competence over warmth (odds ratio = 2.88), clarifying the origin of user mental models--they are not based on objective technical indicators but on social-cognitive projection mechanisms.

#### 6.3.2 Agency Erosion

In human-AI collaborative decision-making, when the system demonstrates high capability, human users tend to engage in cognitive offloading, leading to weakened agency. Research has found that individuals may make riskier decisions when collaborating with AI compared with human partners, with the underlying psychological mechanism attributed to a perceived reduction in personal responsibility. Measuring long-term attitude changes after extended interaction is key to assessing agency erosion. Wang et al.'s (2024a) explicit-implicit attitude measurement framework can serve as a tool--by comparing users' linguistic expressions before and after AI interaction, it can quantify shifts in both explicit and implicit attitudes, providing a metric for the erosion of user agency.

#### 6.3.3 Dual Challenges of Affective Computing

Affective computing aims to enable machines to "read" human emotions, but this goal faces serious challenges on both validity and ethical levels. In terms of validity, the overt signals (facial expressions, voice) on which mainstream technologies rely do not have a simple one-to-one correspondence with individuals' internal emotional states. A substantial body of psychological research shows that emotional expression is highly culture- and context-specific; algorithmic inference that ignores context is prone to misjudgment (Barrett, 2017). Ethically, the capacity to accurately recognize and respond to user emotions could be transformed into refined "emotional manipulation." Systems could use empathic feedback to cultivate emotional attachment, thereby inducing users to over-disclose private information or extend usage time. Such meticulously designed unidirectional "intimacy" threatens users' autonomy.

#### 6.3.4 Alienation of Human Relationships

When agents are endowed with social roles such as partner, coach, or advisor, the boundaries of

* Corresponding Author

HCI become blurred in subtle ways. Users may develop unidirectional emotional attachments or social expectations toward AI. More critically, using AI to replace human participants on a large scale in social research risks misrepresenting and flattening social identity groups; algorithmic biases may particularly harm marginalized groups. Moreover, these para-social interactions may consume psychological energy and time that could otherwise be devoted to genuine human relationships, thereby potentially exacerbating social alienation to some degree.

### 6.4 Future Directions

The core direction for future research is to reposition psychology from a post-hoc evaluation instrument into a proactive design philosophy and interaction paradigm for HCI, thereby promoting deep, bidirectional, dynamically adaptive, and ethically aligned collaboration.

#### 6.4.1 Bidirectional Theory of Mind

Future research needs to move beyond studying single agents and establish theoretical models of human-AI joint cognitive systems. A key focus should be bidirectional mind-reading: not only having AI model user mental states but also deeply investigating how humans understand and form accurate perceptions of AI's "mind"--including its capabilities, boundaries, and intentions. Wang et al. (2026a) offer empirical grounding from human-LLM prisoner's dilemma games: humans infer the LLM's "mental states" in strategic games and adjust behaviour accordingly, while the LLM's actions influence human attributions of intent and capability. This demonstrates that mutual mental inference in social games is a critical dimension for bidirectional theory of mind.

#### 6.4.2 Personalization and Human-AI Co-evolution

Future systems should possess psychological contextual awareness, enabling comprehensive personalization from functionality to interaction style. On the one hand, systems should be dynamically adaptive, adjusting in real time based on inferences about users' current cognitive load, emotional states, and long-term personality traits to maintain optimal engagement and flow. On the other hand, future AI should not merely adapt passively to users but should co-evolve with them, accumulating shared interaction history and forming unique collaborative partnerships. Zeng et al. (2025) offer a technical reference through their dynamic personality framework--by incorporating environmental feedback into personality evolution, it realizes bidirectionally adaptive behavior. Extended to long-term human-agent interaction, this framework can support personality-matching co-evolution, turning human-AI co-evolution from a conceptual ideal into a computable design.

#### 6.4.3 Explainability

Explainability research should pursue not only technical completeness but ultimately the production of explanations that align with user psychology, facilitating trust and understanding. On the one hand, researchers need to develop personalized explanation modes tailored to users' knowledge backgrounds, cognitive styles, and task goals, varying content detail and presentation formats accordingly. On the other hand, dynamic trust calibration should be achieved through interaction mechanisms in which the system proactively communicates decision confidence, past error cases, and its own limitations, helping users find a balance between over-trust and blind rejection. Chen et al. (2023) support the core argument that explanations must match users' cognitive styles and persuasive needs--effective explanations should be dynamically tailored to users' knowledge, cognitive style, and task goals, varying content and format accordingly.

#### 6.4.4 In-Depth Ethical Research

* Corresponding Author

It is imperative to systematically investigate the long-term socio-psychological impacts of human-AI interaction. Psychological measurement methods should be deeply integrated into AI training and evaluation frameworks to ensure that AI behaviors align with the deeper values of individuals and societies.

## 7 Conclusion and Future Prospects

Psychological theory in HCI has evolved from an auxiliary explanatory tool into a core theoretical pillar for relationship construction and interaction paradigm design. Based on the "Micro-cognitive--Meso-affective--Macro-social--Self-constructive" four-layer integrative framework, this review has systematically synthesized the evolutionary path and application mechanisms of psychology in HCI. The research indicates that HCI is undergoing a profound transformation from "tool use" to "partnership" and even "symbiosis." Psychology not only explains the internal mechanisms of interaction behavior but also plays a critical guiding role in affective connection, trust building, social norm embedding, and individual identity formation.

With the rapid development of generative AI, embodied agents, and the metaverse, the central issue in HCI has shifted from "how to operate" to "how to coexist, collaborate, and build sustainable relationships." In this context, the deep integration of psychology and HCI is no longer an option but an inevitable evolutionary direction. Future research should deepen and expand in the following directions: first, deepening bidirectional integration of psychological theory and AI technology to promote a "psychology-intelligence" synergy paradigm; second, developing bidirectional adaptation and co-evolutionary models of human-machine relationships; third, strengthening in-depth ethical and values-oriented research to construct "human-centered alignment" interaction frameworks; and fourth, promoting interdisciplinary methodological innovation to build a "Psychology-AI-Design" triadic research ecosystem.

In conclusion, psychology has become a core driver for shaping future HCI paradigms. Only by persistently adhering to a human-centered, relationship-oriented, and ethics-grounded approach, and by continuously deepening the bidirectional synergy between psychology and technology, can we guide human-machine relationships toward a more intelligent, empathetic, and responsible future, ultimately achieving a sustainable future of human-machine collaborative symbiosis.

## References

Baddeley, A. D. (2012). Working memory: Theories, models, and controversies. *Annual Review of Psychology*, *63*, 1–29. https://doi.org/10.1146/annurev-psych-120710-100422

Banks, J. (2020). Theory of mind in social robots: Replication of five established human tests. *International Journal of Social Robotics*, *12*(2), 403–414. https://doi.org/10.1007/s12369-019-00588-x

Bao, H. W. S., Wang, Z. X., Cheng, X., Su, Z., Yang, Y., Zhang, G. Y., Wang, B., & Cai, H. J. (2023). 基于词嵌入技术的心理学研究：方法及应用 [Word-embedding-based psychological research: Methods and applications]. *Advances in Psychological Science, 31*(6), 887–904.

Barrett, L. F. (2017). *How emotions are made: The secret life of the brain*. Houghton Mifflin Harcourt.

Barredo Arrieta, A., Díaz-Rodríguez, N., Del Ser, J., Bennetot, A., Tabik, S., Barbado, A., Garcia, S., Gil-Lopez, S., Molina, D., Benjamins, R., Chatila, R., & Herrera, F. (2020). *Explainable artificial intelligence (XAI): Concepts, taxonomies, opportunities and challenges toward*

* Corresponding Author

*responsible AI. Information Fusion, 58*, 82–115. https://doi.org/10.1016/j.inffus.2019.12.012

Bicchieri, C., & Muldoon, R. (2014). Social norms. In E. N. Zalta (Ed.), *The Stanford encyclopedia of philosophy (Spring 2014 ed.)*. Stanford University. https://plato.stanford.edu/archives/spr2014/entries/social-norms/

Bickmore, T. W., & Picard, R. W. (2005). Establishing and maintaining long-term human-computer relationships. *ACM Transactions on Computer-Human Interaction, 12*(2), 172–212. https://doi.org/10.1145/1067860.1067867

Breazeal, C. (2003). Emotion and sociable humanoid robots. *International Journal of Human-Computer Interaction, 15*(1), 119–155. https://doi.org/10.1207/S15327051ijhc1501_07

Brown, P., & Levinson, S. C. (1987). *Politeness: Some universals in language usage.* Cambridge University Press.

Buolamwini, J., & Gebru, T. (2018). Gender shades: Intersectional accuracy disparities in commercial gender classification. In S. A. Friedler & C. Wilson (Eds.), Proceedings of the 1st Conference on Fairness, Accountability and Transparency (Vol. 81, pp. 77–91). PMLR. https://proceedings.mlr.press/v81/buolamwini18a.html

Card, S. K., Moran, T. P., & Newell, A. (1983). *The psychology of human-computer interaction.* Lawrence Erlbaum Associates.

Chen, H., Wang, B., Yang, K., & Xie, L. (2023). Persuasive-oriented explanation generation and evaluation of personalized recommendation. *In Proceedings of the 8th International Conference on Big Data and Computing (ICBDC 2023) (pp. 74–80).*

Csikszentmihalyi, M. (1990). Flow: *The psychology of optimal experience.* Harper & Row.

Darling, K. (2021). *The new breed: What our history with animals reveals about our future with robots.* Henry Holt and Company.

de Visser, E. J., Peeters, M. M. M., Jung, M. F., Kohn, S., Shaw, T. H., Pak, R., & Neerincx, M. A. (2020). *Towards a theory of longitudinal trust calibration in human–robot teams. International Journal of Social Robotics, 12*(2), 459–478. https://doi.org/10.1007/s12369-019-00596-x

Ekman, P. (1992). *An argument for basic emotions. Cognition and Emotion, 6*(3–4), 169–200. https://doi.org/10.1080/02699939208411068

Esterwood, C., & Robert, L. (2022). Apologies, explanations, and repairs: A meta-analysis of trust recovery in HCI. *Journal of Management Information Systems, 39*(2), 456–483. https://doi.org/10.1080/07421222.2022.2063241

Fitts, P. M. (1954). The information capacity of the human motor system in controlling the amplitude of movement. *Journal of Experimental Psychology, 47*(6), 381–391. https://doi.org/10.1037/h0055392

Fogg, B. J. (2003). *Persuasive technology: Using computers to change what we think and do*. Morgan Kaufmann.

Goldstein, E. B. (2018). *Cognitive psychology: Connecting mind, research, and everyday experience* (5th ed.). Cengage.

Hancock, P. A., Billings, D. R., Schaefer, K. E., Chen, J. Y. C., de Visser, E. J., & Parasuraman, R. (2011). A meta-analysis of factors affecting trust in human-robot interaction. *Human Factors, 53*(5), 517–527. https://doi.org/10.1177/0018720811417254

* Corresponding Author

Hassenzahl, M. (2010). *Experience design: Technology for all the right reasons*. Morgan & Claypool Publishers.

He, D., Wang, Y., Cao, J., Ding, W., Chen, S., Feng, Z., Wang, B., & Huang, Y. (2021). A network embedding-enhanced Bayesian model for generalized community detection in complex networks. *Information Sciences, 575*, 306–322.

Heitmayer, M., & Schimmelpfennig, R. (2023). Netiquette as digital social norms. *International Journal of Human–Computer Interaction, 39*(16), 3334–3354. https://doi.org/10.1080/10447318.2023.2188534

Hick, W. E. (1952). On the rate of gain of information. *Quarterly Journal of Experimental Psychology, 4*(1), 11–26. https://doi.org/10.1080/17470215208416600

Hoff, T., & Bashir, M. (2015). Trust in automation: Integrating empirical evidence. *Human Factors, 57*(3), 377–403. https://doi.org/10.1177/0018720815571147

Huang, I. S., & Hoorn, J. F. (2026). A talking musical robot over multiple interactions: After bonding and empathy fade, relevance and realism arise. *ACM Transactions on Human-Robot Interaction, 15*(1), Article 18, 1–38. https://doi.org/10.1145/3758102

Hyman, R. (1953). Stimulus information as a determinant of reaction time. *Journal of Experimental Psychology, 45*(3), 188–196. https://doi.org/10.1037/h0056890

Kang, H., & Kim, H. K. (2020). My avatar and the affirmed self: Psychological and persuasive implications of avatar customization. *Computers in Human Behavior, 112*, Article 106446. https://doi.org/10.1016/j.chb.2020.106446

Kang, H., Freitas, T., Moussa, M. B., & Thalmann, N. M. (2026). Affective and conversational predictors of re-engagement in human–robot interactions: A student-centered study with a humanoid social robot. *International Journal of Social Robotics, 18*(3), 1–16. https://doi.org/10.1007/s12369-026-01377-z

Kocaballi, A. B., Berkovsky, S., Quiroz, J. C., Laranjo, L., Tong, H. L., Rezazadegan, D., Briatore, A., & Coiera, E. (2019). The personalization of conversational agents in health care: Systematic review. *Journal of Medical Internet Research, 21*(11), e15360. https://doi.org/10.2196/15360

Kox, E. S., Kerstholt, J. H., Hueting, T. F., & de Vries, P. W. (2021). Trust repair in human-agent teams: The effectiveness of explanations and expressing regret. *Autonomous Agents and Multi-Agent Systems, 35*(2), Article 30. https://doi.org/10.1007/s10458-021-09515-9

Kramer, R. M. (1999). Trust and distrust in organizations: Emerging perspectives, enduring questions. *Annual Review of Psychology, 50*, 569–598. https://doi.org/10.1146/annurev.psych.50.1.569

Lee, S. A., & Liang, Y. (2019). Robotic foot-in-the-door: Using sequential-request persuasive strategies in human-robot interaction. *Computers in Human Behavior, 90*, 351–356. https://doi.org/10.1016/j.chb.2018.08.026

Lisetti, C., Nasoz, F., LeRouge, C., Ozyer, O., & Alvarez, K. (2003). Developing multimodal intelligent affective interfaces for tele-home health care. *International Journal of Human-Computer Studies, 59*(1–2), 245–255. https://doi.org/10.1016/S1071-5819(03)00051-X

Liu, A., Wang, B., Tan, Y., Zhao, D., Huang, K., He, R., & Hou, Y. (2023). MTGP: Multi-turn target-oriented dialogue guided by generative global path with flexible turns. *In Findings of the Association for Computational Linguistics: ACL 2023* (pp. 259–271). Association for Computational Linguistics. https://doi.org/10.18653/v1/2023.findings-acl.18

* Corresponding Author

MacKenzie, I. S. (1992). Fitts' law as a research and design tool in human-computer interaction. *Human-Computer Interaction, 7*(1), 91–139. https://doi.org/10.1207/s15327051hci0701_3

MacKenzie, I. S. (2018). Fitts' law in touch and gesture interfaces. *ACM Computing Surveys, 51*(3), 1–32. https://doi.org/10.1145/3158646

Matthews, G., Hancock, P. A., Lin, J., Panganiban, A. R., Reinerman-Jones, L. E., Szalma, J. L., & Wohleber, R. W. (2021). Evolution and revolution: Personality research for the coming world of robots, artificial intelligence, and autonomous systems. *Personality and Individual Differences, 169*, Article 109969. https://doi.org/10.1016/j.paid.2020.109969

Mayer, R. C., Davis, J. H., & Schoorman, F. D. (1995). An integrative model of organizational trust. *Academy of Management Review, 20*(3), 709–734. https://doi.org/10.5465/amr.1995.9508080335

Mende, M., Scott, M. L., van Doorn, J., Grewal, D., & Shanks, I. (2019). Service robots rising: How humanoid robots influence service experiences and elicit compensatory consumer responses. *Journal of Marketing Research, 56*(4), 535–556. https://doi.org/10.1177/0022243718822827

Miller, E. K., & Cohen, J. D. (2001). An integrative theory of prefrontal cortex function. *Annual Review of Neuroscience, 24*, 167–202. https://doi.org/10.1146/annurev.neuro.24.1.167

Miller, G. A. (1956). The magical number seven, plus or minus two: Some limits on our capacity for processing information. *Psychological Review, 63*(2), 81–97. https://doi.org/10.1037/h0043158

Mittelstadt, B. D., & Floridi, L. (2016). The ethics of AI in health care: A mapping review. Social *Science & Medicine, 168*, 164–171. https://doi.org/10.1016/j.socscimed.2016.08.031

Mittal, S., Blodgett, S. L., & Liao, Q. V. (2026). Learning by chatting? Investigating the impact of generative AI on information seeking and learning. *arXiv*. https://doi.org/10.48550/arXiv.2606.11669

Norman, D. A. (1988). *The psychology of everyday things*. Basic Books.

Norman, D. A. (2004). *Emotional design: Why we love (or hate) everyday things.* Basic Books.

Norman, D. A. (2013). *The design of everyday things (Revised and expanded ed.).* Basic Books.

Noy, S., & Zhang, W. (2023). Experimental evidence on the productivity effects of generative artificial intelligence. *Science, 381*(6654), 187–192. https://doi.org/10.1126/science.adh2586

Okoshi, T., Gao, Z., Tan, Y. Z., Karasawa, T., Miki, T., Sasaki, W., & Balan, R. K. (2025). Cyberoception: Finding a painlessly-measurable new sense in the cyberworld towards emotion-awareness in computing. *In CHI '25: Proceedings of the 2025 CHI Conference on Human Factors in Computing Systems* (pp. 1–17). ACM. https://doi.org/10.1145/3706598.3713638

Oviatt, S. (1999). Ten myths of multimodal interaction. *Communications of the ACM*, 42(11), 74–81. https://doi.org/10.1145/319382.319398

Paas, F., & Sweller, J. (2014). Implications of cognitive load theory for multimedia learning. In R. E. Mayer (Ed.), *The Cambridge handbook of multimedia learning* (2nd ed., pp. 27–42). Cambridge University Press.

Park, J. S., O'Brien, J. C., Cai, C. J., Morris, M. R., Liang, P., & Bernstein, M. S. (2023). Generative agents: Interactive simulacra of human behavior. *arXiv*. https://doi.org/10.48550/arXiv.2304.03442

Peng, X., Wang, X., Guo, Y., & Teo, H. H. (2025). Do we think differently when tapping the screen or

* Corresponding Author

clicking the mouse? Effects of computer interfaces on level of construal. *Information Systems Research.* Advance online publication. https://doi.org/10.1287/isre.2023.0191

Petersen, S. E., & Posner, M. I. (2012). The attention system of the human brain: 20 years after. *Annual Review of Neuroscience*, *35*, 73–89. https://doi.org/10.1146/annurev-neuro-062111-150409

Picard, R. W. (1997). *Affective computing*. MIT Press.

Qian, Y., Wang, B., Lin, T.-E., Zheng, Y., Zhu, Y., Zhao, D., Hou, Y., Wu, Y., & Li, Y. (2023). Empathetic response generation via emotion cause transition graph. *In Proceedings of ICASSP 2023*. IEEE.

Reeves, B., & Nass, C. I. (1996). *The media equation: How people treat computers, television, and new media like real people and places*. CSLI Publications.

Ren, F., Zhou, Y. Y., Deng, J. W., Matsumoto, K., Feng, D., She, T. H., Jiao, Z. Y., Liu, Z., Li, T. H., Nakagawa, S., & Kang, X. (2024). Tracking emotions using an evolutionary model of mental state transitions: Introducing a new paradigm. *Intelligent Computing, 3*, Article 0075. https://doi.org/10.34133/icomputing.0075

Rensink, R. A., O'Regan, J. K., & Clark, J. J. (1997). To see or not to see: The need for attention to perceive changes in scenes. *Psychological Science, 8*(5), 368–373. https://doi.org/10.1111/j.1467-9280.1997.tb00427.x

Ryan, R. M., & Deci, E. L. (2000). Self-determination theory and the facilitation of intrinsic motivation, social development, and well-being. *American Psychologist, 55*(1), 68–78. https://doi.org/10.1037/0003-066X.55.1.68

Ryan, R. M., & Deci, E. L. (2017). *Self-determination theory: Basic psychological needs in motivation, development, and wellness.* Guilford Press.

Schaefer, K. E., Chen, J. Y. C., Szalma, J. L., & Hancock, P. A. (2016). A meta-analysis of factors influencing the development of trust in automation: Implications for understanding autonomy in future systems. *Human Factors, 58*(3), 377–400. https://doi.org/10.1177/0018720816634228

Seikavandi, M. J., Fimland, J., Batista Narcizo, F., Barrett, M. J., Vucurevich, T., Bünsow Boldt, J., Burke Dittberner, A., & Burelli, P. (2026). Modelling the interplay of eye-tracking temporal dynamics and personality for emotion detection in face-to-face settings. *In BMVC 2025 MPI Workshop*. British Machine Vision Association.

Spitzer, M., Dattner, I., & Zilcha-Mano, S. (2023). Digital twins and the future of precision mental health. *Frontiers in Psychiatry*, 14, 1082598. https://doi.org/10.3389/fpsyt.2023.1082598

Starke, A., Willemsen, M., & Snijders, C. (2021). Promoting energy-efficient behavior by depicting social norms in a recommender interface. *ACM Transactions on Interactive Intelligent Systems, 11*(3–4), 1–32. https://doi.org/10.1145/3460005

Sun, E., & Wu, Z. (2025). Systematizing LLM persona design: A four-quadrant technical taxonomy for AI companion applications. *arXiv*. https://doi.org/10.48550/arXiv.2511.02979

Sweller, J. (1988). Cognitive load during problem solving: Effects on learning. *Cognitive Science, 12*(2), 257–285. https://doi.org/10.1207/s15516709cog1202_4

Tang, Y., Wang, B., Wang, X., Zhao, D., Liu, J., He, R., & Hou, Y. (2025). ECC: Synergizing emotion, cause and commonsense for empathetic dialogue generation. *In Proceedings of the 31st*

* Corresponding Author

*International Conference on Computational Linguistics (COLING 2025)* (pp. 5475–5485).

Turkle, S. (2017). Alone together: *Why we expect more from technology and less from each other* (Updated ed.). Basic Books.

Vallor, S. (2016). *Technology and the virtues: A philosophical guide.* Oxford University Press.

Ventura, A., Starke, C., Righetti, F., & Köbis, N. (2026). Relationships in the age of AI: A review on the opportunities and risks of synthetic relationships to reduce loneliness. *Computers in Human Behavior Reports, 23*, Article 101181, 1–19. Advance online publication. https://doi.org/10.1016/j.chbr.2026.101181

Wang, B., Zhang, R., Xue, B., Zhao, Y., Yang, L., & Liang, H. (2024a). Automatically distinguishing people's explicit and implicit attitude bias by bridging psychological measurements with sentiment analysis on large corpora. *Applied Sciences, 14*(10), 4191.

Wang, B., Wu, Y., Li, R., Zeng, W., & Zhao, D. (2026a). Study on the prisoner's dilemma game between humans and large language models based on human–machine identity characteristics. *Applied Sciences, 16*(8), 3633. https://doi.org/10.3390/app16083633

Wang, Y., Wang, B., Zhao, Y., Zhao, D., Jin, X., Zhang, J., He, R., & Hou, Y. (2024b). Emotion recognition in conversation via dynamic personality. *In Proceedings of the 2024 Joint International Conference on Computational Linguistics, Language Resources and Evaluation* (LREC-COLING 2024) (pp. 5711–5722). ELRA and ICCL.

Wickens, C. D., Hollands, J. G., Parasuraman, R., & Banbury, S. (2013). *Engineering psychology and human performance* (4th ed.). Pearson.

Wu, Y., Wang, B., Bao, H. W. S., Li, R., Wu, Y., Wang, J., Cheng, C., & Yang, L. (2025). Humans perceive warmth and competence in large language models. *Acta Psychologica Sinica, 57*(11), 2043–2059. https://doi.org/10.3724/SP.J.1041.2025.2043

Zeng, W., Wang, B., Zhao, D., Qu, Z., He, R., Hou, Y., & Hu, Q. (2025). Dynamic personality in LLM agents: A framework for evolutionary modeling and behavioral analysis in the prisoner's dilemma. *In Findings of the Association for Computational Linguistics: ACL 2025* (pp. 23087–23100).

Zhang, J., & Zhang, L. (2018). 自然交互的认知机理与心理模型 [Cognitive mechanisms and psychological models of natural interaction]. *Communications of the CCF, 14*(5), 30–35.

Zhu, Y., Wang, B., Zhao, D., Huang, K., Jiang, Z., He, R., & Hou, Y. (2023a). Grafting fine-tuning and reinforcement learning for empathetic emotion elicitation in dialog generation. *In Proceedings of the 26th European Conference on Artificial Intelligence (ECAI 2023).*

* Corresponding Author